# Thermal Radiation (TR) mode: A Deployment Perspective for 5G NR

Haneet Kour, *Student Member, IEEE*, Rakesh Kumar Jha, Senior *Member,IEEE, Sanjeev Jain, Member IEEE,* Shubha Jain, *Student Member, IEEE*

## I. INTRODUCTION

The 5G New Radio (NR) technology is under standardization process by 3GPP to provide outline for a new radio interface for the next generation of cellular networks. The aim of the 5G networks include not only to provide enhanced capacity, coverage but also support advanced services such as enhanced mobile broadband (eMBB), Ultra-Reliable Low Latency Communication (URLLC), massive Machine Type Communication (mMTC). Key features of NR include Ultra lean carrier design to minimize the power consumption by limiting the "always-on" signal transmissions and to reduce interference in the neighboring cells (Parkvall et al., 2017). Another feature is the use of massive number of antennas for transmission as well as reception of signals. This rise in the number of antennas to provide a greater coverage brings about various challenges and impact in the system.

With the increase in investigations in the mmWave frequencies, there is a need to investigate the health hazards they have on human body and the environment at large. This paper intends to provide an insight into the harmful impacts of Radio Frequency (RF) fields. The radiation metric to study the RF impact for far field is power density and for near field is Specific Absorption Rate (SAR). These are the two main EM radiation metrics to find out the exposure due to uplink and downlink phenomenon in mobile communications. Mobile communication systems are addressed particularly to discuss the Electromagnetic (EM) Radiation impact as smart phones are used in close proximity to the body. A proposal in the form of "Thermal Radiation" (TR) mode is given to reduce the radiations emitted from a mobile phone. The performance of the proposed mode is validated from the results by achieving reduced power density, complexity and exposure ratio.

## II. IMPACT OF 5G NR IN REAL TIME DEPLOYMENT

The adverse impacts caused by EM waves are studied by standard organizations such as ICNIRP (see ICNIRP 2018) and IEEE C95. The restrictions presented in these documents are based on scientific data and are studied by measuring the effects in terms of electric and magnetic fields. Rising power density levels disturbs the process of photosynthesis in plants and trees, which leads to their poisoning and hinders growth and physiological processes (Tang, C. et al., 2018). As birds and animals can detect magnetic stimulus, rising EM radiations has lead to a decline in the population of birds such as sparrow (Cucurachi et al., 2013). Studies in literature have validated a reduction in growth rate in animals due to EM radiations (Cucurachi et al., 2013). Also, extreme weather conditions are on the rise every year due to rise in emission of harmful gases such as $CO_2$ (179 Megatons Carbon dioxide emission (Mt $CO_2 e$) because of increased greenhouse effect (SMART 2020).

In humans, initial symptoms such as headache, eye, skin problems and in worst case traces of carcinogenicity and other diseases such as Alzeimer's and Parkinson's disease can also occur (see R.S Kshetrimayum, 2008, T. Wu et al., 2015). Fig. 1 is a diagrammatic representation of the findings from R.S Kshetrimayum, 2008, T. Wu et al., 2015, M.A. Jamshed et al., 2019 and WHO International EMF Project for diseases and effects on human beings. The effect on animals, birds and insects are findings from Cucurachi et al., 2013 and the destruction of nature by electromagnetic pollution are findings from Tang, C. et al., 2018. The increase in $CO_2$ footprint due to radio frequency waves is based on findings from SMART 2020. Various works discuss the new opportunities for reducing the EM exposure and planning the future networks in a way so that the EMF (Electromagnetic Field) limits and constraints are abided (L.Chiaraviglio et al., 2018). There is a requirement to update the safety guidelines and the EMF evaluation framework keeping in view the impact of wireless communication industry on EMF exposure (M.A. Jamshed et al., 2019).

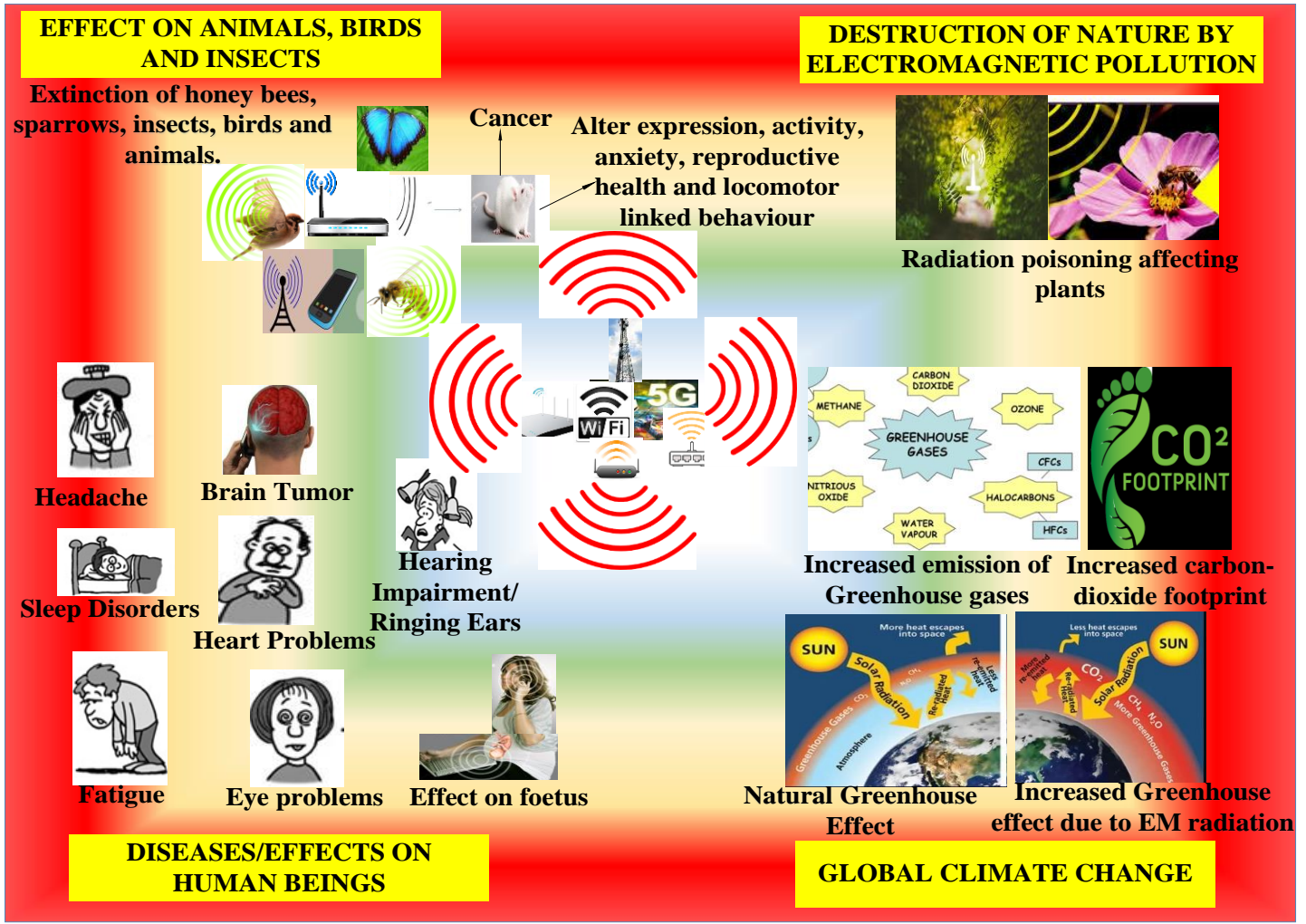

FIG 1. General Architecture corroborating EM radiation impact.

III. ELECTROMAGNETIC RADIATION REDUCTION TECHNIQUES

For the future generation networks, the heterogeneity in the network increases as we move towards Ultra Dense Networks (UDN's) with incorporation of newer technologies. There will be a significant role of EM radiations that will be emitted with these networks and the effects produced by them. Below we discuss some techniques that can be incorporated to reduce this radiation impact (Sambo, et al 2014).

**Massive MIMO (Multiple Input Multiple Output) :** It consists of a single base station or transmitter deployed with large number of small antennas on it that increase the spatial diversity by transmitting parallel streams of information channels . Each antenna deployed is a low power antenna. Higher the count of the antennas on the base station, lower is the power level required for each antenna. With this technique the total power transmitted by the base station significantly decreases. **Electromagnetic Shielding**: This technique incorporates the use of conductive materials or ferrite materials to block the electromagnetic field between the mobile device and human head. It reduces SAR for the body, called as SAR Shielding /RF Shielding. **Beamforming:** Beamforming has the advantage of decreasing the transmission power required in a system without compromising with the quality of service. As the transmitted beam is directed towards a desired receiver the electromagnetic radiation towards the undesired users is reduced along with reducing the interference caused to them, hence promoting green communication.

Another reduction technique is **Coordinated multipoint (CoMP)** .In this technique the transmission and the reception points coordinate dynamically to have enhanced coverage, increase the throughput and spectral efficiency of a system. **Spectrum Sharing:** Sharing the available radio resource is necessary and it can largely help in efficient utilization of the EM spectrum. Sharing of the licensed, unlicensed and TV white spaces by the secondary users of the spectrum can improve the overall radiation exposure produced in a network. It has still been found that the expected percentage of $CO_2$ footprint has not reduced and is expected to be rising considerably with the years to come. This calls for more enhanced techniques or proposals to be introduced for the upcoming 5G New Radio scenario.

## IV. HALF DUPLEX RADIO: ROADMAP FOR THERMAL RADIATION MODE

A large amount of interference caused in the UDN's increases the requirement of self-interference cancellation techniques such as NOMA (Non-Orthogonal multiple access) schemes and so on. Various challenges in a full duplex system include UL-to-DL (Uplink-to-Downlink) interference among different users in a single cell scenario and a complex version of it in a multicell scenario where there can be DL-to-DL (Downlink-to-Downlink) interference between user equipments, UL-to-UL interference at the base stations. They have an adverse impact on the overall performance of a system. In Fig. 2 a proposal has been presented to reduce the EM radiation impact and improve the SINR (Signal-to-noise-plus-Interference Ratio) for mobile communication systems. The scenario presents conventional approach and proposed approach. There is consideration of 50 users in the AM, and in the TR mode region 30 users are assumed to be in AM and 20 in TR mode. Non-stationary and wideband channels are assumed considering Rayleigh fading distribution (R.K Jha et al 2020).

Catering to the issue of "always on" signals in the current mobile communication systems, the TR mode switches the communication mode from full duplex to half-duplex. When the "TR mode" is switched ON in the network, only the downlink information transfer signals remain active and the uplink information transfer signals are disabled. The reference signals required for downlink information transfer and to maintain a stable connection with the base station only remain active i.e. the reference signals are ON only when there is data to be transmitted and the base station detection signal is active (Fig. 2).

The proposed methodology for a device transitioning in TR mode is based on adaptive switching operation. The switching occurs when the mobile phone does not have adequate signal strength to support applications requiring high bandwidth. We consider a mobile device communicating with "always-on" signals in a cell. The received signal strength changes in accordance to the channel conditions between the base station and the mobile user. When the received signal strength degrades below a threshold, the device cannot support applications requiring high data rate. We propose adapting switching in the mode of the device from "Active" to "TR". With this mode the device only supports applications requiring low data rate, such as voice calls and regular text messages. All the high bandwidth requiring applications cannot be supported until the signal strength improves. Whenever the channel conditions are favorable and the received signal strength is greater than the threshold, the device automatically transitions to AM thus supporting all the applications with "Always-on" signals.

This mode improves the SINR of the entire network. With respect to near field communication, the thermal heating produced in the device is reduced which in turn reduces the SAR. With respect to the far field communication the power density in the network reduces, hence both the factors collaboratively help in reducing the EM radiation exposure produced in the network.

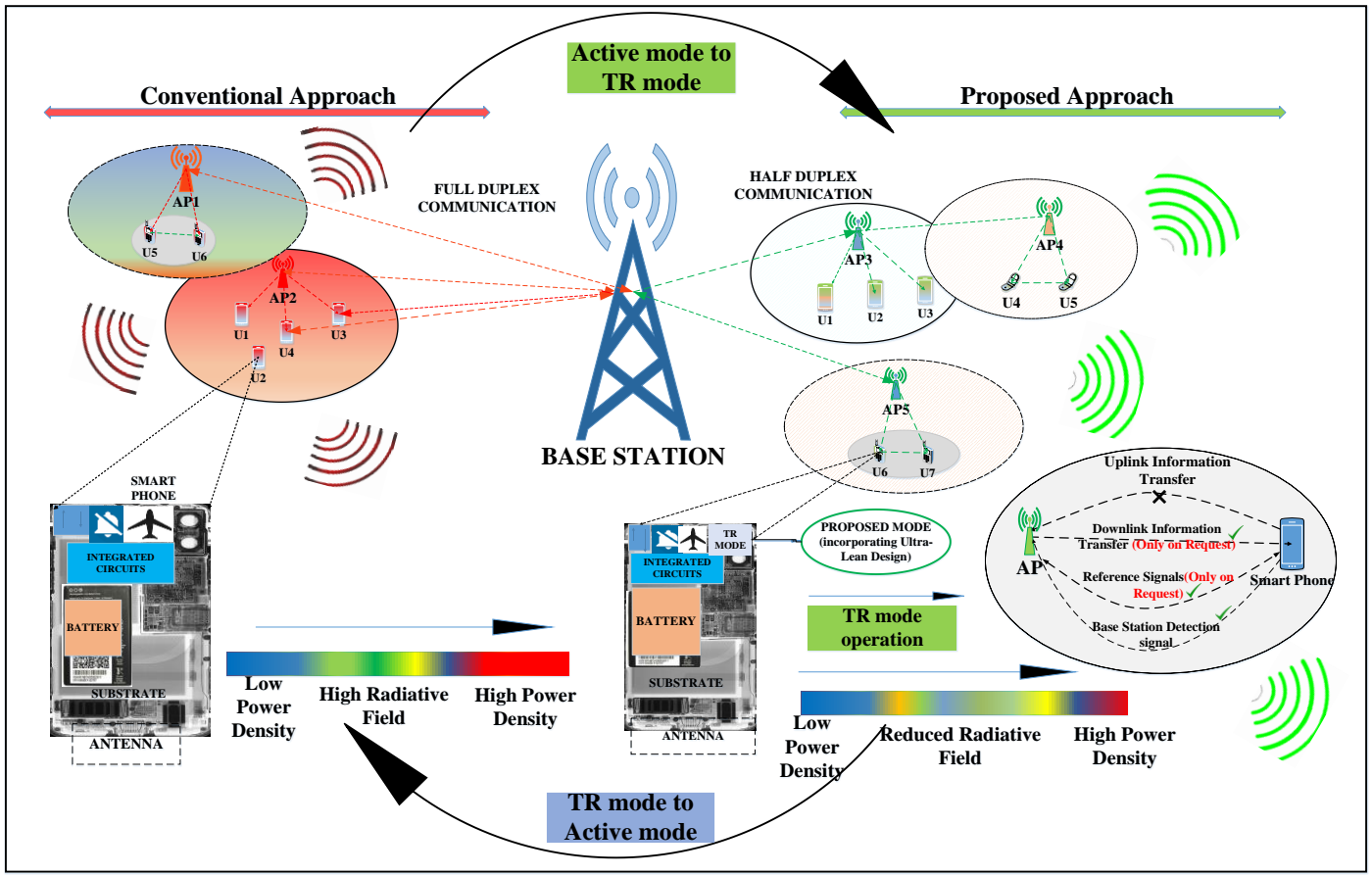

FIG 2. Proposed Thermal Radiation (TR) mode.

## V. ADAPTIVE FRAME STRUCTURE DESIGN FOR TR MODE IN 5G NR

The waveforms supported in 5G NR are similar to LTE which include CP-OFDM (UL/DL) with DFT-s-OFDM (UL). There is flexible numerology support in 5G NR with 15 KHz × 2n subcarrier spacing. The proposed frame structure for 5G NR is depicted in Fig. 3 keeping the LTE (Long-Term Evolution) frame format as basis, along with the supported duplex schemes for spectrum allocation. One radio frame consists of 10 subframes each of 1ms duration. The duplex schemes supported are similar to LTE i.e. Frequency division duplex (FDD) and Time division duplex (TDD). As depicted for the full duplex frame type, both the UL and DL frames are of 10ms duration each and are transmitted simultaneously as they are separated by different UL and DL frequencies. For the TDD frame type, the transmission frequency remains the same and the multiplexing takes place in time domain. For the FDD frame type, on activation of the TR mode, the mobile only receives downlink information and does not transmit in the uplink. A frequency switching subframe has been added in the uplink frame which on activation of TR mode changes its state from "0" to "1" i.e. it stops the uplink transmission. On activation of TR mode in the handset slot 1 is active i.e. the frequency switching occurs otherwise the uplink frame continues its transmission and slot 0 is active. For the proposed TDD frame type a subframe has been added in the previously called "special subframe" now called "superframe (TR Mode)". On activation of the TR mode the added subframe goes in the "hold" mode i.e. it avoids the switching of the subframe from DL to UL and stops all the uplink transmissions in the TDD frame. When the TR mode is deactivated it is released and the TDD frame continues its transmissions in the conventional way. The hold/release mode are allocated one slot each in a subframe for the proposed mode.

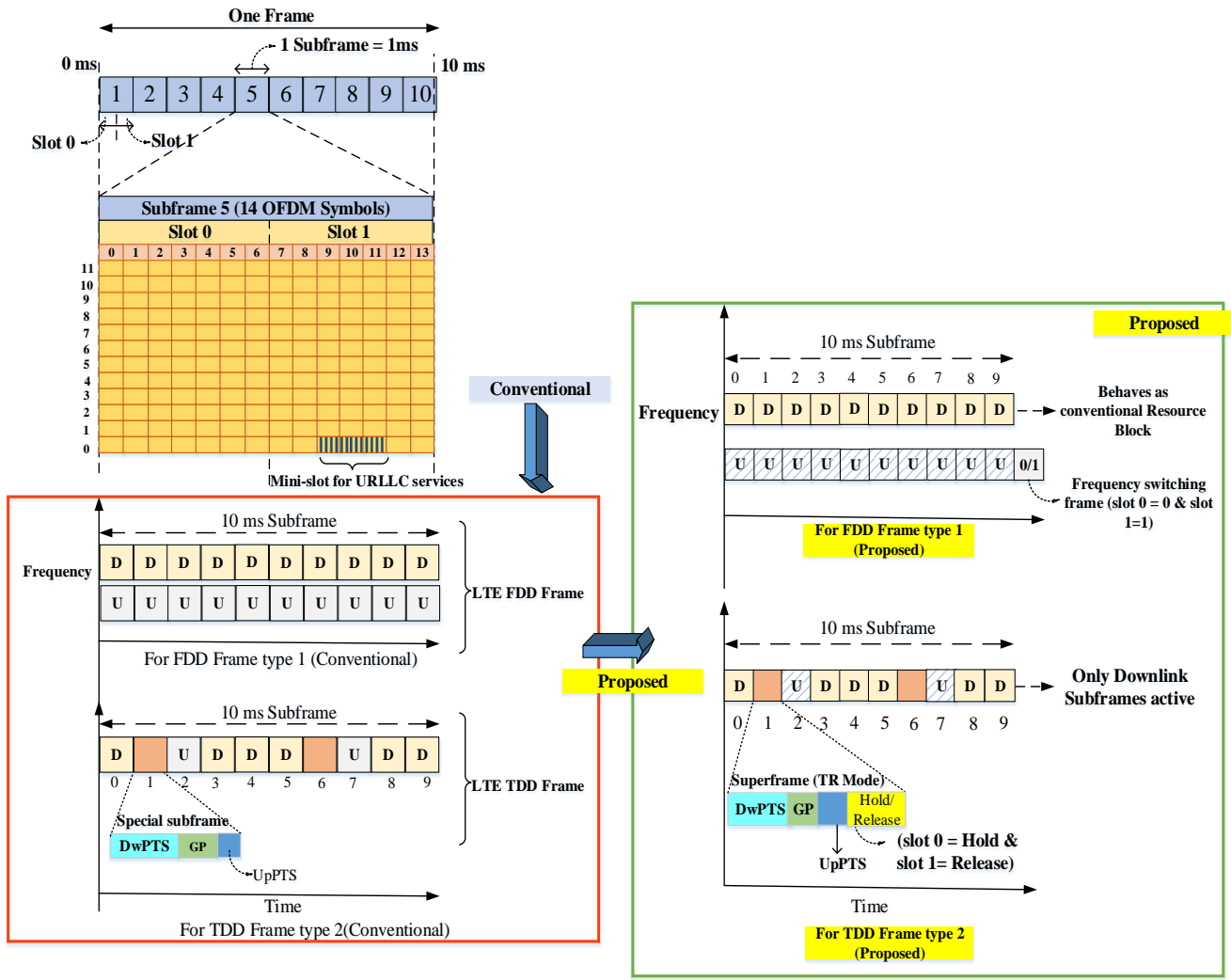

FIG 3. Frame format for 5G NR (FDD and TDD Frame type).

## VI. RADIO RESOURCE CONTROL (RRC) STATE TRANSITION HANDLING FOR TR MODE

According to the standardization for 5G NR there are three protocol states that exist in the RRC state machine: RRC_Idle, RRC_Inactive and RRC_Connected. RRC_Idle is optimized for lesser consumption of power as well as resources in the network. RRC_Connected is for high activity of the user equipment. RRC_Inactive is a state that reduces/lightens the transition procedure .We propose a RRC_EE (i.e. RRC_Energy Efficient) state for the proposed TR mode which is a low activity state in the transition model.

In the proposed state transition model for TR mode, the state machine consists of the RRC Idle, RRC Active, RRC Inactive and a novel RRC EE (Energy Efficient) as depicted in Fig. 4. The proposed state i.e. RRC_EE is the state corresponding to the TR mode i.e. a low activity state. The characteristic of RRC EE is that 5GC-NG-RAN connection is kept for the user equipment only for downlink information transfer and not for uplink information transfer. This reduces the overhead signaling in the air interface.

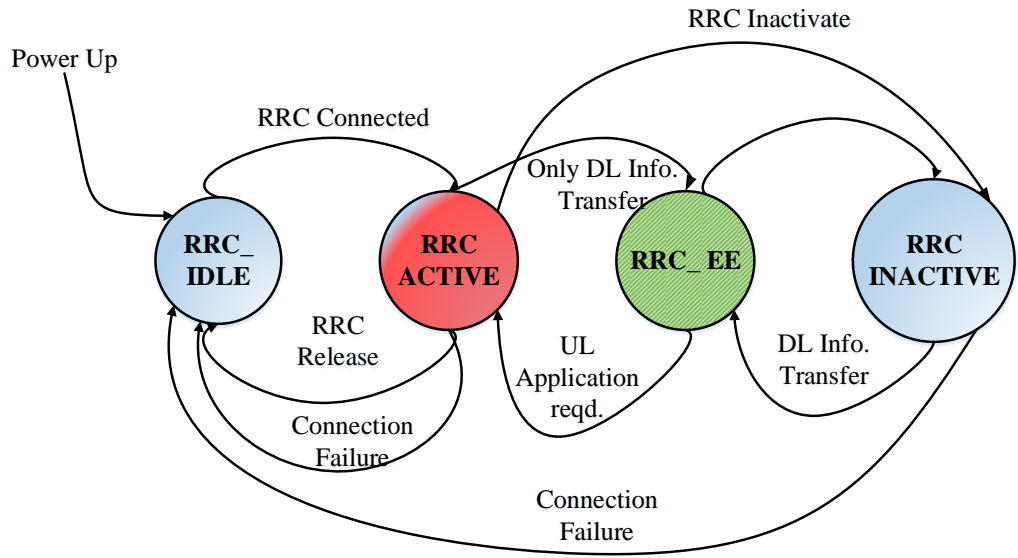

FIG 4. Potential state transition model for TR mode in 5G.

The transitions between the Idle, Connected and the Inactive state is expected to follow the LTE procedure. The state transition to RRC EE is initiated when the user equipment is in TR mode. Whenever there is uplink information transfer to take place the mode is changed to active and AM communication resumes. This state has been configured for quick transitions and to incorporate URLLC services.

## VII. PERFORMANCE ANALYSIS

The performance of the proposed mode is presented in terms of reduced power density, Exposure Ratio (ER) and complexity of the system. As we move on to the higher generations of wireless communications the power density levels also increase correspondingly in the atmosphere. Fig. 5 depicts power density as a function of rising generations of wireless communication with highest value being obtained for the 5$^{th}$ Generation (IEEE Std. C95, 2006). The comparison of power density is made with TR mode. As the TR mode operation suspends the uplink transfer of information there is less overhead signaling involved. This decreases the transmitted power from each device and hence the overall power density in the network. The initial values and the new values obtained are given in the graph in Fig. 5.

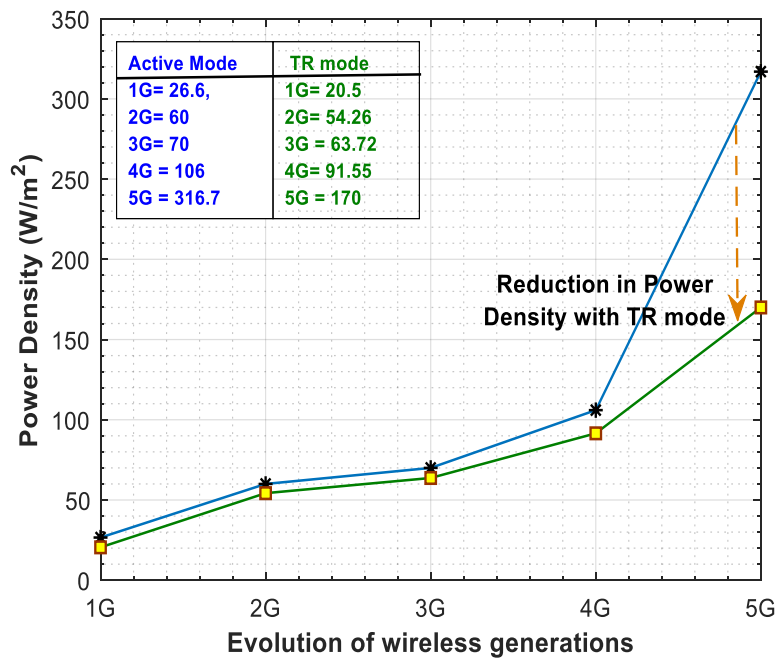

FIG 5. Evolution of wireless generations v/s Power density.

TR mode in a device reduces the ER obtained which is a ratio of the electric field produced by a device to the reference level of electric field or Maximum permissible exposure allowed in that area. The comparison of ER is made for 1G to 5G generations with ICNIRP and IEEE C95 standards as depicted in Table 1. It is evident that TR mode reduces ER in both the scenarios for ICNIRP as well as IEEE C95.

TABLE 1: Comparison of Exposure Ratio (ER) for Wireless Generations with ICNIRP and IEEE C95 Standards.

| Wireless Generations | ICNIRP | | Wireless Generations | IEEE C95 | | Observation |
| --- | --- | --- | --- | --- | --- | --- |
| | Exposure Ratio (ER) | | | Exposure Ratio (ER) | | |
| | Active Mode (AM) | Thermal Radiation (TR) mode | | Active Mode (AM) | Thermal Radiation (TR) mode | It is observed from the comparison of ER for 1G to 5G communication network that incorporation of TR mode decreases the ER in each of the generations substantially. |
| 1G | 0.2403 | 0.211 | 1G | 0.1848 | 0.1622 | |
| 2G | 0.361 | 0.3432 | 2G | 0.277 | 0.264 | |
| 3G | 0.41 | 0.3719 | 3G | 0.31 | 0.286 | |
| 4G | 0.481 | 0.4458 | 4G | 0.369 | 0.3429 | |
| 5G | 0.83 | 0.6075 | 5G | 0.6379 | 0.4673 | |

With the suspension of the uplink information transfer signals, the EM radiations produced from the device on activation of TR mode are also limited. Hence, the interference produced due to each device operating in TR mode reduces with suspension of some signals temporarily. A 3D pattern for rising ER for Active Mode and TR mode is obtained. The ER values obtained rise as we move from 1G to 5G and are evidently lower for TR mode in comparison to AM (fig. 6).

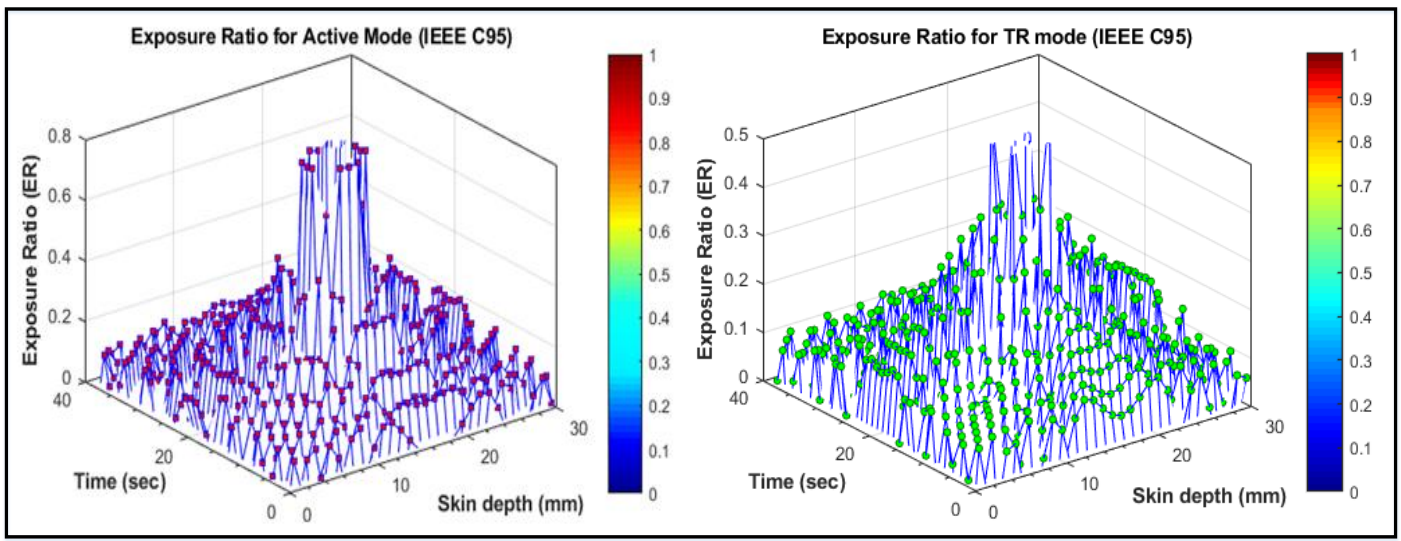

FIG 6. 3D patterns for Exposure Ratio (ER) rise in Active mode and TR mode for Wireless Generations.

As we have considered a non-stationary and wideband channel there is fluctuation in the received Signal-to-Noise ratio due to fading, multipath etc. Outage probability has been obtained for devices communicating in AM and TR mode. The devices communicating in TR mode produce less interference in the network that improves the SNR associated. This helps in achieving low EM radiation exposure without compromising the target data rate. The outage probability graph in Fig. 7 denotes the low SNR region as high EM radiation zone, and when the device transitions to TR mode there is improved SNR obtained with less EM radiation emission.

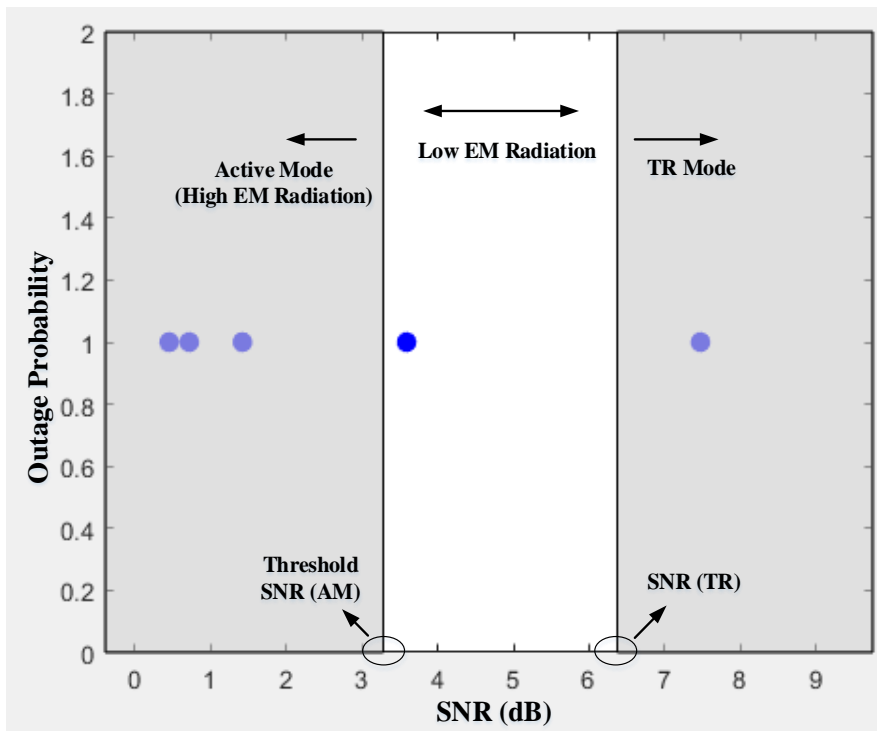

Fig. 7 Outage probability v/s SNR for Active Mode and TR mode.

The reduction in interference caused due to adjacent cells in TR mode redues the complexity of the entire system. The graph in Fig. 8 depicts a comparison of the complexity in the network as a function of the interference power for active devices in AM and TR mode. It is evident from the graph that there is reduced complexity in the network for TR mode in comparison to AM.

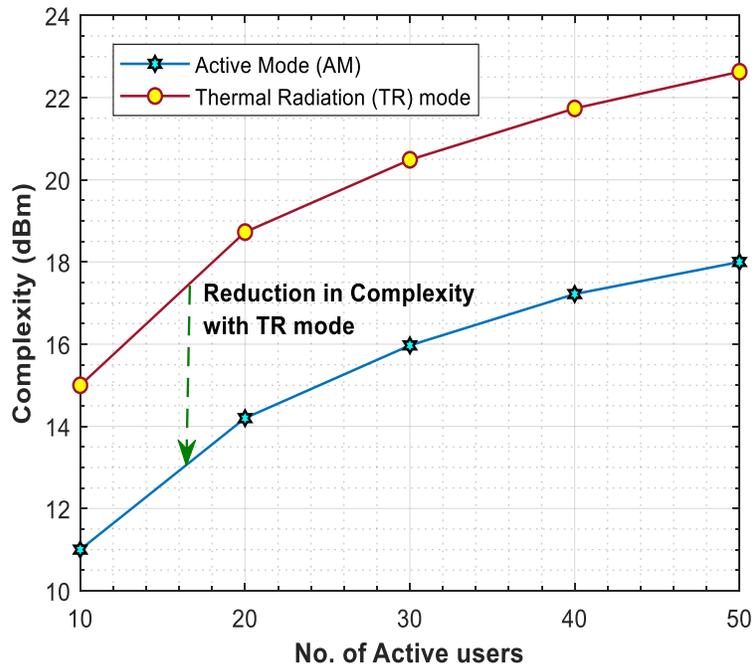

FIG 8. Complexity v/s No. of active users.

## VIII. CONCLUSION

This article aims to provide an insight into the 5G NR interface standard that will be operating at a very wide spectrum range utilizing high frequency bands to incorporate a large number of devices and high bandwidth demanding applications. The concern of high EM radiation exposure that arises with the growing wireless communications is addressed in the article. Rising EM radiations are responsible for increasing power density levels in the environment. A proposal has been given for the mobile communications with a proposed mode (TR mode) to reduce the radiation levels in the atmosphere and also improve the biological safety of the human exposure. For the proposal, a frame structure design and RRC state handling procedures have also been presented for 5G NR. The performance of the proposed mode can be seen in terms of reduction in power density, complexity and exposure ratio produced.

ACKNOWLEDGEMENT

The authors gratefully acknowledge the support provided by 5G and IoT Lab, DoECE and TBIC, Shri Mata Vaishno Devi University, Katra, Jammu.

BIOGRAPHIES

HANEET KOUR (hani.kpds@gmail.com) received the B.E degree in ECE engineering from Jammu University, Jammu and Kashmir, India, in 2015 and M.Tech degree from SMVDU, India in 2017. She is currently pursuing Ph.D degree at SMVDU (Shri Mata Vaishno Devi University), India. Her research interest includes the emerging technologies of 5G wireless communication network. Currently she is doing her research on Power Optimization in next generation networks. She is working on MATLAB tools for Wireless Communication.

Dr. RAKESH K JHA (jharakesh.45@gmail.com) is currently an Associate Professor in School of ECE Engineering, SMVDU Jammu and Kashmir, India. He has published several SCI Journals Papers including IEEE Transactions, IEEE Journal and IEEE Magazine. His area of interest is Wireless communication, Optical Fiber Communication, Security issues. He has received APAN fellowship in 2011, 2012, 2017 and 2018. He is also a member of ACM, CSI, many patents and has more than 2900 Citations in his credit.

PROF. SANJEEV JAIN (dr_sanjeevjain@yahoo.com) obtained his Post Graduate Degree in CS Engineering from IIT, Delhi. He received his Doctorate Degree in CS Engineering and has over 24 years' experience in teaching and research. He has served as Director, Madhav Institute of Technology and Science (MITS), Gwalior. Presently, he is working at IIITDM Jabalpur, Madhya Pradesh, India. Professor Jain has the credit of making significant contribution to R & D in the area of Image Processing and Mobile Adhoc Network.

SHUBHA JAIN received the bachelor's degree in electronics and telecommunications from SGSITS, Indore, India. She has completed her master's degree in telecommunications from the University of Maryland, College Park, MA, USA. Currently she is working with Amazon, USA.